\documentclass[preprint,prc,aps,tightenlines,floatfix,showpacs,superscriptaddress]{revtex4-1}
\usepackage{graphics}
\usepackage{amsmath}
\usepackage{bm}

\begin{document}

\title{Proton scattering observables from Skyrme-Hartree-Fock densities}

\author{S. Karataglidis} 
\email{stevenka@uj.ac.za}
\affiliation{Department of Physics, University of Johannesburg, P.O. Box 524, 
Auckland Park, 2006,  South Africa}

\author{K. R. Henninger}
\email{g06h0128@campus.ru.ac.za}
\affiliation{Department of Physics and Electronics, Rhodes
  University, P.O. Box 94, Grahamstown, 6140, South Africa}

\author{W. A. Richter}
\email{richter@sun.ac.za}
\affiliation{Department of Physics, University of the Western Cape, 
  Private Bag X17, Bellville 7530, South Africa}

\author{K. Amos}
\email{amos@unimelb.edu.au}
\affiliation{School of Physics, The University of Melbourne, 
Victoria 3010, Australia}

\begin{abstract}
  Proton and neutron densities from Skyrme-Hartree-Fock (SHF)
  calculations are used to generate non-local ($g$-folding)
  proton-nucleus optical potentials. They are formed by folding the
  densities with realistic nucleon-nucleon interactions.  The
  potentials are then used to calculate differential cross sections
  and spin observables for proton scattering. Good agreement with data
  has been found, supporting those found previously when using SHF
  charge densities in analyses of electron scattering data.  That
  agreement was improved by use of (shell model) occupation numbers to
  constrain the HF iterations. That, in part, is also the case with
  analyses of proton scattering data. The $g$-folding method is
  extended to exotic nuclei by including data for neutron-rich
  $sd$-shell nuclei from the inverse kinematics of scattering from
  hydrogen.
\end{abstract}
\date{\today}
\maketitle
\section{Introduction}

Nucleon-nucleus ($NA$) scattering probes the matter density of the
nucleus, and of particular interest are the separate density
distributions of protons and neutrons. The differences between the two
densities become increasingly important as one moves away from the
valley of stability towards the drip lines, with the emergence of
structures such as halos and skins. Those structures influence $NA$
scattering, whence the problem becomes one of how to extract such
information from measurement. For detailed information on the matter
(proton and/or neutron) density of the target nucleus one requires a
microscopic description of the interaction between the projectile and
the target, necessitating also a microscopic (nucleon-based)
description of the nucleus.

For radioactive nuclei, experiments have been done in inverse
kinematics with a beam of nuclei incident on a hydrogen target. Thus
the calculation of differential cross sections and spin observables
requires some microscopic model structure for the exotic nucleus,
which determines the ground-state densities. The adequacy of the model
assumed, for stable nuclei, can be checked against information
obtained from the elastic scattering of electrons.  The electron
scattering form factors are measures of the charge and current
densities of the nucleus. While electron scattering data for exotic
nuclei are not presently available, the SCRIT experiment
\cite{Su05,*Su09} and the electron-ion collider ELISe at FAIR
\cite{Si05} will measure such form factors. At present, one must rely
on model predictions for the matter densities and test those against
proton scattering data.

Proton and neutron ground-state densities have been determined for a
wide range of nuclei from Skyrme-Hartree-Fock (SHF) calculations, and
extensive comparisons for the charge densities have been made with
available data from electron scattering \cite{Ri03}.  The good
agreement generally found reflects on the adequacy of the calculated
matter densities, which are used as input for the calculation of
proton elastic scattering observables in this work.  Analyses of
proton scattering data from $^{208}$Pb using SHF models have been made
\cite{Ka02}, whereby the neutron skin thickness in $^{208}$Pb was
determined. One corollary of the latter was the observation that
analyses of zero momentum transfer data are not enough to elicit
sufficient information on the densities of nuclei. Data taken at
finite momentum transfer are also required. For analyses of proton
scattering data to have credibility, as tests of the underlying
structures assumed for the target nuclei, one must have a model of
scattering for which there are no parameters to be fitted to the data
being analysed. Only then may one evaluate the models used to specify
the nuclear structure without ambiguity.
 
The matter densities obtained from the assumed structure models
(shell, SHF, and SHF with shell-model occupancies) have been used with
realistic nucleon-nucleon interactions in a folding model, requiring
no \textit{a posteriori} fitting \cite{Am00}, to specify optical
potentials for the elastic scattering of protons with energies in the
range $25 - 200$~MeV.  Those potentials are then used to make
predictions for differential cross sections and spin observables for
proton scattering, a procedure that has been applied extensively
\cite{Am00}.  Comparisons are then made to available proton scattering
data. When using that folding model with the SHF densities, those allowed for
the recognition of a signature for exotic structures in neutron-rich
isotopes in the reaction cross sections \cite{Am06}. The present work
extends that earlier work to consider the effects on the elastic
scattering observables.

\section{The Skyrme-Hartree-Fock density calculations}

Charge-density distributions and the associated nuclear radii have
been calculated with the Hartree-Fock method for comparison with
available data from electron scattering \cite{Ri03}. Two forms of the
Skyrme interaction have been used for the calculations, the so-called
SKX$_{\text{csb}}$ \cite{Br00} and SKM* \cite{Ba82} interactions. The
SKX$_{\text{csb}}$ Hamiltonian is based on the SKX Hamiltonian
\cite{Br98} with a charge-symmetry-breaking (CSB) interaction, added
to account for nuclear displacement energies \cite{Br00}. The charge
densities from SKX and SKX$_{\text{csb}}$ are essentially identical,
so the SKX$_{\text{csb}}$ results will be referred to as "SKX". The
SKX and SKM* results for the charge densities are very similar, the
main difference being that the interior density is about 5\% higher
with SKM*, with marginally better overall agreement with experiment
for SKM*.  Generally good agreement between theory and experiment has
been achieved in extensive comparisons of measured nuclear
charge-density distributions with calculated values for $p$-shell,
$sd$-shell, and $fp$-shell nuclei and some selected magic and
semi-magic nuclei up to Pb. The extent of the agreement is further
improved by constraining the Hartree-Fock equations to use occupation
numbers from large-basis shell model calculations.  Somewhat larger
deviations are observed for lighter nuclei, which may imply the
inadequacy of the mean-field approximation \cite{Ri03}. The good
agreement with experiment for electron scattering data is a
justification for using the proton and neutron radial wave functions
as input for the optical model calculations.

For the purposes of the present study, we have used both the basic SHF
(denoted SHF henceforth) densities and those obtained from the SHF
constrained by use of the shell model occupation numbers (denoted as
SHF-SM).  (Those numbers are listed in Table~\ref{table-occ} for the
isotopes considered herein.)
\begin{table}
\begin{ruledtabular}
\caption{Shell-model occupation numbers, as used in the SHF-SM calculations, for the given orbits. \\
\centerline{(Nomenclature: $4=0d_{5/2}, 5=0d_{3/2}, 6=1s_{1/2}, 7=0f_{7/2}, 
8=0f_{5/2}, 9=1p_{3/2}, 10=1p_{1/2}$)}}
\label{table-occ}
\begin{tabular}{ccccc}
Z & Isotope & Orbits & Proton occupancies & Neutron occupancies \\
\hline
14 & $^{28}$Si & 4, 5, 6    & $4.623, 0.673, 0.704$ & $4.623, 0.673, 0.704$  \\
16 & $^{32}$S &  4, 5, 6    & $5.421, 1.161, 1.418$ & $5.421, 1.161, 1.418$  \\ 
16 & $^{34}$S &  4, 5, 6    & $5.607, 0.736, 1.657$ & $5.674, 2.484, 1.752$  \\ 
16 & $^{36}$S & 4, 5, 6     & $5.869, 0.237, 1.894$ & $6.000, 4.000, 2.000$ \\
16 & $^{38}$S & 4, 5, 6     & $5.766, 0.762, 1.472$ & \\
   &           & 7, 8, 9, 10    &    & $1.636, 0.088, 0.242, 0.034$ \\
 16 & $^{40}$S & 4, 5, 6   & $5.727, 1.302, 0.972$ & \\
   &           & 7, 8, 9, 10   &    & $3.350, 0.156, 0.443, 0.052$ \\
18 & $^{36}$Ar & 4, 5, 6 & $5.761, 2.500, 1.740$ & $5.761, 2.500, 1.740$ \\
18 & $^{38}$Ar & 4, 5, 6 & $5.946, 2.109, 1.945$ & $ 6.000, 4.000, 2.000$ \\
18 & $^{40}$Ar & 4, 5, 6    & $5.920, 2.200, 1.880$ &                     \\
   &          & 7, 8, 9, 10 &           & $1.750, 0.060, 0.160, 0.030$  \\
 18 & $^{44}$Ar & 4, 5, 6 & $5.840, 2.659, 1.491$ & \\
   &         & 7, 8, 9, 10 &          & $5.359, 0.193, 0.397, 0.051$ \\
20 & $^{40}$Ca &  5, 6, 7, 9  & $3.090, 1.800, 0.990, 0.120$ &  
$3.090, 1.800, 0.990, 0.120 $ \\ 
20 & $^{42}$Ca & 7, 8, 9, 10 & & $1.812, 0.058, 0.110, 0.021$ \\
20 & $^{44}$Ca & 7, 8, 9, 10 & & $3.674, 0.126, 0.171, 0.029$ \\
20 & $^{48}$Ca &  5, 6, 7, 9  & $3.550, 1.630, 0.750, 0.070$  & 
$3.830, 1.960, 7.910, 0.300$  \\
\end{tabular}
\end{ruledtabular}
\end{table}
For the $sd$-shell, the USD interaction of Wildenthal and Brown
\cite{Br88} as well as the USDB interaction of Brown and Richter
\cite{Br06}, in the case of the neutron-rich nuclei, were used. For
those nuclei in which the neutrons extend to the $fp$-shell the
\textit{sdpf}-U interaction of Nowacki and Poves \cite{No09} was used
in the $sdpf$-model space. No spuriosity is admitted into the wave
functions as the protons are restricted to be solely within the
$sd$-shell. All calculations of the occupation numbers were done using
the NuShell shell-model code \cite{Nu10}. For comparison, we have also used shell model wave functions
where indicated for the lighter isotopes, and used a simple packed shell model in which the lowest orbits are
filled, with no configuration mixing.

\section{Calculation of proton scattering observables}

To calculate microscopically the differential cross sections and spin
observables for $NA$ scattering, one generally begins with an
effective nucleon-nucleon ($NN$) interaction. That interaction is
folded with the ground-state density of the target nucleus to obtain
the microscopic optical potential from which the observables are
obtained.  Herein, we utilise the Melbourne approach, which is
described in detail in a review \cite{Am00}; we give a brief outline
to highlight the important aspects of this model below.

To obtain a credible effective $NN$ interaction one usually starts
with the $g$ matrices of the free $NN$ interaction. Those $g$ matrices
are solutions of the Brueckner-Bethe-Goldstone (BBG) equations for
infinite nuclear matter, \textit{viz.}
\begin{equation}
  g\left(  \mathbf{q}',  \mathbf{q};   \mathbf{K}  \right)  =  V\left(
  \mathbf{q}',   \mathbf{q}  \right)   +  \int   V\left(  \mathbf{q}',
  \mathbf{k}'  \right)  \frac{  Q\left( \mathbf{k}',  \mathbf{K};  k_f
  \right) }{  \left[ E\left( \mathbf{k}, \mathbf{K}  \right) - E\left(
  \mathbf{k}',  \mathbf{K} \right) \right]  } \,  g\left( \mathbf{k}',
  \mathbf{q}; \mathbf{K} \right) \, d\mathbf{k}',
\end{equation}
where $Q$ is a Pauli blocking  operator, and effects of the mean field
are  incorporated into  the auxiliary  potentials entering  the energy
denominator. The  centre-of-mass and Fermi momenta are  denoted by $K$
and $k_f$, respectively.

The $g$ matrices obtained from the BBG equations are mapped to a
coordinate-space representation to give an effective $g$ matrix
($g_{\text{eff}}$) which is complex, energy- and density-dependent.
When that effective interaction is folded with a reasonable
microscopic model structure of the target for the nucleus, one obtains
the $NA$ optical potential of the form,
\begin{align}
  U\left( \mathbf{r}, \mathbf{r}'; E \right) & = \delta\left(
  \mathbf{r} - \mathbf{r}' \right) \sum_i n_i \int
  \varphi^{\ast}_{i}(\mathbf{s}) g_D\left( \mathbf{r}, \mathbf{s}; E
  \right) \varphi_i(\mathbf{s}) \, d\mathbf{s} + \sum_i n_i \varphi^{\ast}_i(
  \mathbf{r})   g_E\left( \mathbf{r}, \mathbf{r}'; E \right) \varphi_i(
  \mathbf{r}') \nonumber \\
  & = U_D\left( \mathbf{r}; E \right) \delta\left( \mathbf{r} -
  \mathbf{r}' \right) + U_E\left( \mathbf{r}, \mathbf{r}'; E
  \right),
\end{align}
where the subscripts $D,E$ denote the (local) direct and (nonlocal)
exchange parts of the optical potential, respectively. The sums are
taken over the bound state single-particle orbits for which $n_i$ are
the associated occupation numbers. We use a variant of the DWBA98
program \cite{Ra98} to calculate the optical potential and observables
using SHF single-particle wave functions. The resultant complex,
energy-, and density-dependent $g$-folding optical potential contains
central, two-body spin-orbit, and tensor terms. As there are no parameter
adjustments to fit to the data being described, all model results are predictions.

\section{Comparisons with data}
We present analyses with reference to various isotopes of S, Ar, and
Ca, as listed in Table~\ref{tab-isotopes}, together with references to
available data as used in these analyses.
\begin{table}
\begin{ruledtabular}
  \caption{Isotopes considered for the elastic scattering of
    protons. References for the available data are given.}
\label{tab-isotopes}
\begin{tabular}{lc}
  Nuclei & References \\
  \hline
  $^{28}$Si & \cite{Ka85}, \cite{Hi88,*Lu92} \\
  $^{32}$S, $^{34}$S, $^{36}$S, $^{38}$S, $^{40}$S & \cite{Ka85},
  \cite{Le80}, \cite{Al85}, \cite{Ho90}, \cite{Ke97}, \cite{Ma99} \\
  $^{36}$Ar, $^{38}$Ar, $^{40}$Ar, $^{42}$Ar, $^{44}$Ar & \cite{Sc00}, 
  \cite{Sa82}\\
  $^{40}$Ca, $^{42}$Ca, $^{44}$Ca, $^{46}$Ca, $^{48}$Ca, $^{50}$Ca,
  $^{52}$Ca, $^{54}$Ca &  \cite{Sa82}, \cite{No81}  \\
\end{tabular}
\end{ruledtabular}
\end{table}
In the following diagrams, except when explicitly specified, the results of the calculations 
made using the shell (or packed shell), SHF, and SHF-SM models are denoted by
the solid, dashed, and dot-dashed lines, respectively.

\begin{figure}
\scalebox{0.7}{\includegraphics*{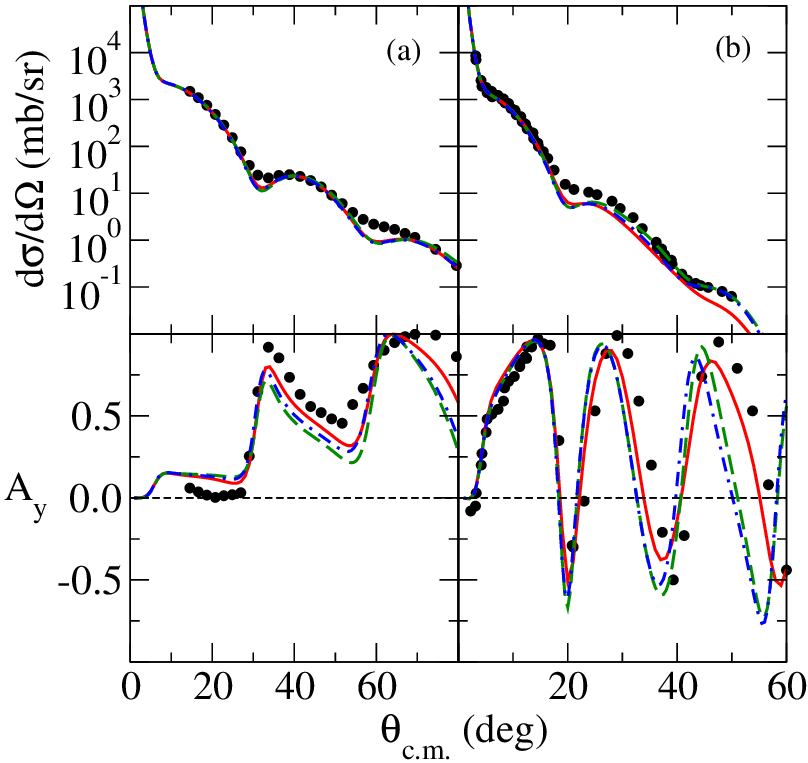}}
\caption{\label{si28-scatt} (Color online.) Differential cross
  sections and analysing powers for the elastic scattering of 65 [(a)]
  and 200~MeV [(b)] protons from $^{28}$Si. The data are compared to
  the results of calculations made using the shell,
  SHF, and SHF-SM models.}
\end{figure}

Before considering the S, Ar, and Ca isotopes, we consider scattering
from $^{28}$Si, as there are many data available for this stable
nucleus. Of the elastic proton scattering data available, we have
chosen to analyse those data taken at 65~MeV \cite{Ka85} and 200~MeV
\cite{Hi88}, as there have been extensive analyses of data for a range
of nuclei at both energies \cite{Do97,*Do98}, resulting in the
effective $NA$ interactions at both energies being
well-established. Fig.~\ref{si28-scatt} presents the differential
cross sections and analysing powers for the elastic scattering
of protons from $^{28}$Si, wherein the data are compared to the
results of the calculations made using wave functions from the shell,
SHF, and the SHF-SM models.  The oscillator length for the harmonic oscillator
single-particle wave functions used in the shell model calculation was
1.85~fm. For 65~MeV scattering there is very little difference in the
predictions for the differential cross sections between the three
models used; they do equally well in describing the data. Differences
emerge in the analysing power, with the shell model result doing best
of all, although all three results describe the shape and magnitudes
well. While the three models do well to describe the forward-angle
differential cross section at 200~MeV, both SHF models do better in
predicting the cross section above $20^{\circ}$.  In the results for the analysing
power at that energy, the shell-model result deviates significantly from the
SHF results above $20^{\circ}$ but are in better agreement with the data.

%
%
\subsection{The S isotopes}
\begin{figure}
\scalebox{0.5}{\includegraphics*{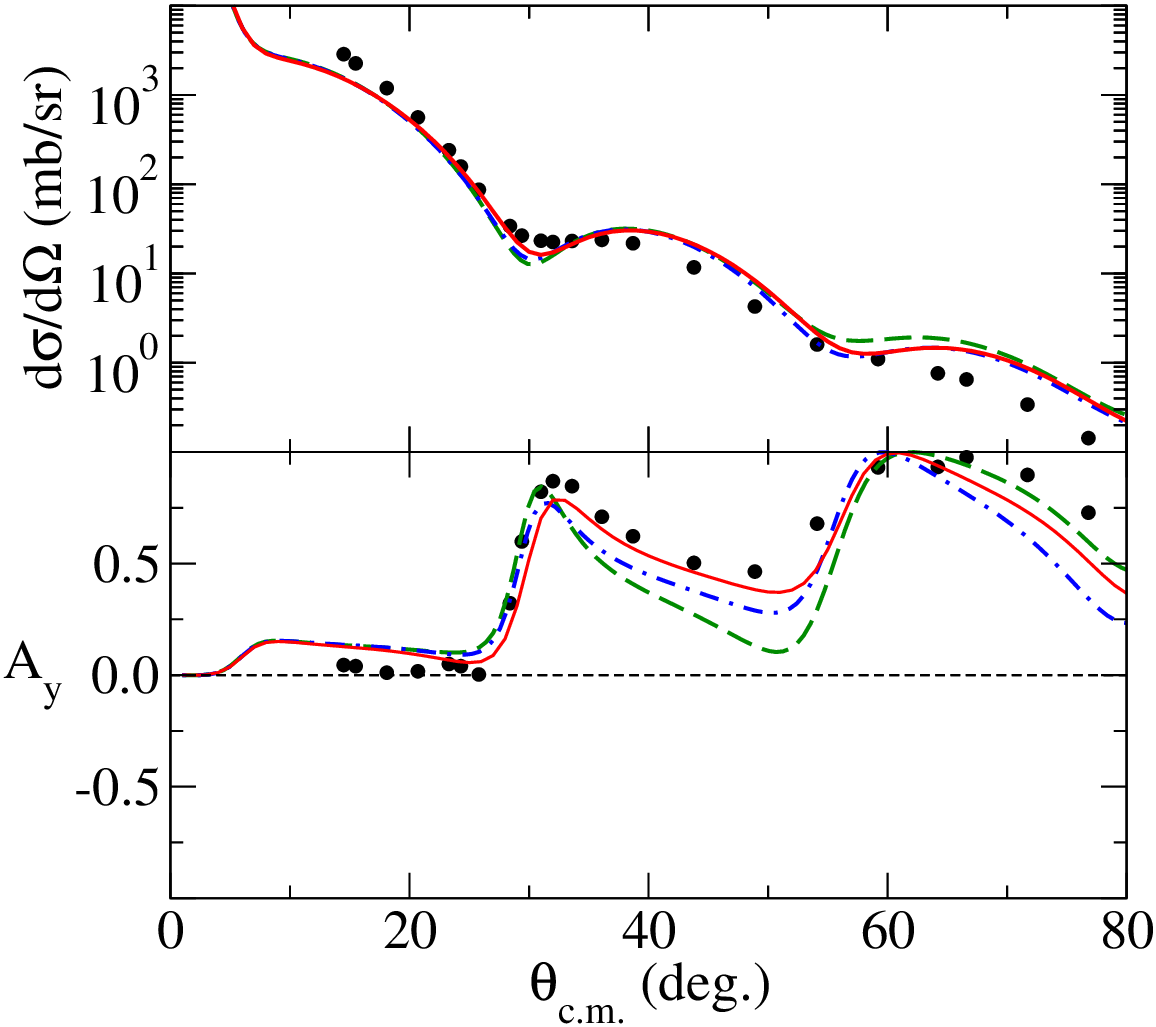}}
\caption{\label{s32-scatt} (Color online.) Differential cross sections
  and analysing powers for the elastic scattering of 65 MeV protons
  from $^{32}$S. The data are compared to the results of calculations
  made using the shell, SHF, and SHF-SM models.}
\end{figure}

\begin{figure}
\scalebox{0.5}{\includegraphics*{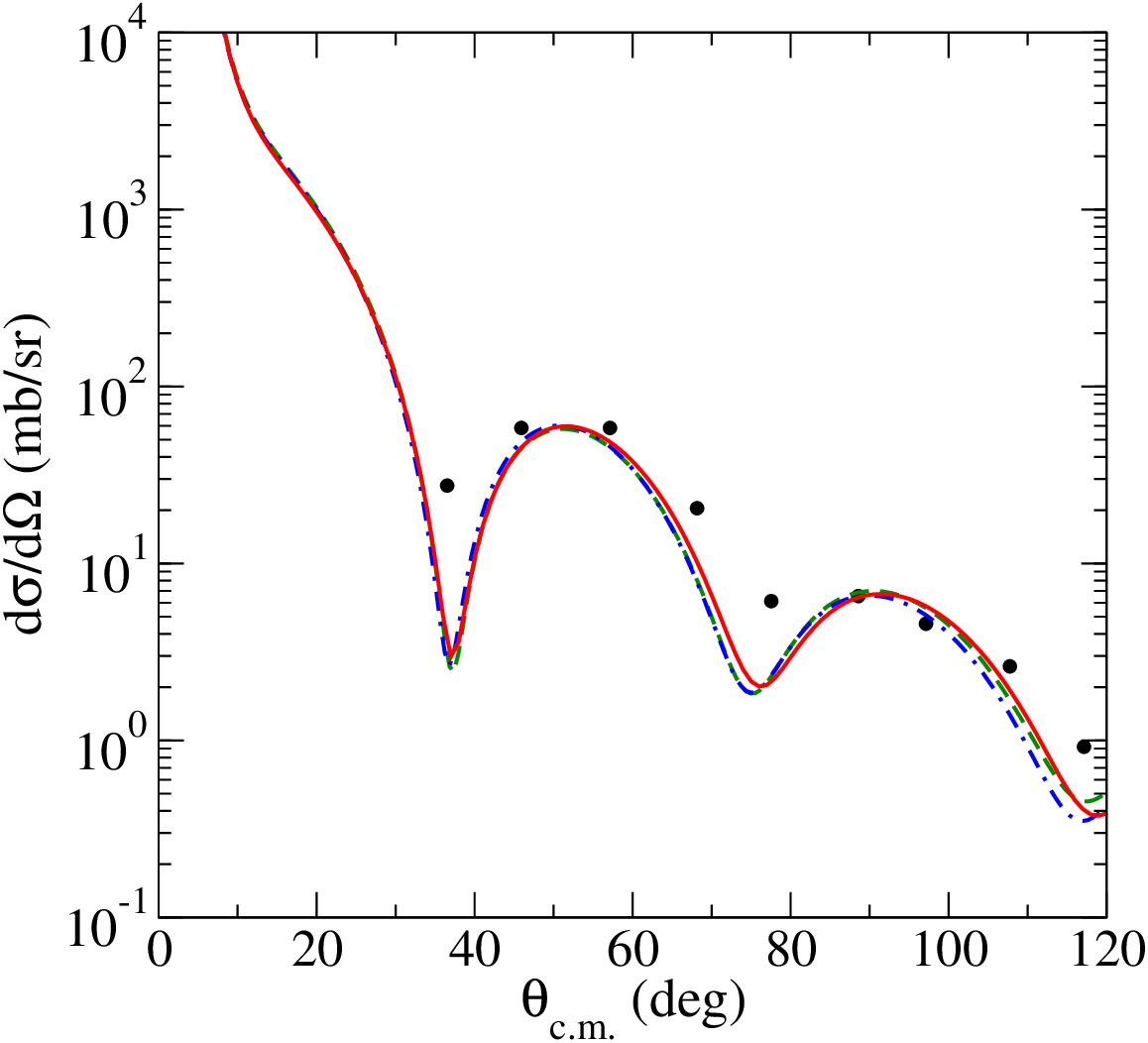}}
\caption{\label{s32_30scatt} (Color online.) As for
  Fig.~\ref{s32-scatt} but for 29.6~MeV protons.}
\end{figure}

Fig.~\ref{s32-scatt} displays the results of calculations made for the
elastic scattering of 65~MeV protons from $^{32}$S.
Therein, the data \cite{Ka85} are compared to the results obtained
from the shell ($b = 1.85$~fm), SHF, and SHF-SM models. All three models
give a reasonable representation of the differential cross-section
data to $60^{\circ}$, but overestimate the differential cross section
at larger angles. The differences between the models are far more
noticeable in the analysing power, for which the shell model result
gives clearly the best agreement with data. Of the two SHF models, the
model with the shell model occupancies defined \textit{a priori} gives
the better agreement. This is of note as $^{32}$S is mid-shell and the
occupancies are expected to play a significant role. We have also
considered the elastic scattering of 29.6~MeV protons from $^{32}$S,
the results of which are shown in Fig.~\ref{s32_30scatt}.
There is agreement between all the models and reasonable agreement with the data \cite{Le80}. 
It is worth noting that the present analysis reproduces the
shape and magnitude well, without the need for \textit{ad hoc} renormalisations
of potentials or the use of the phenomenological $M_n/M_p$ ratios
\cite{Ma99}, which has been shown to be problematic \cite{Am10}.

\begin{figure}
\scalebox{0.5}{\includegraphics*{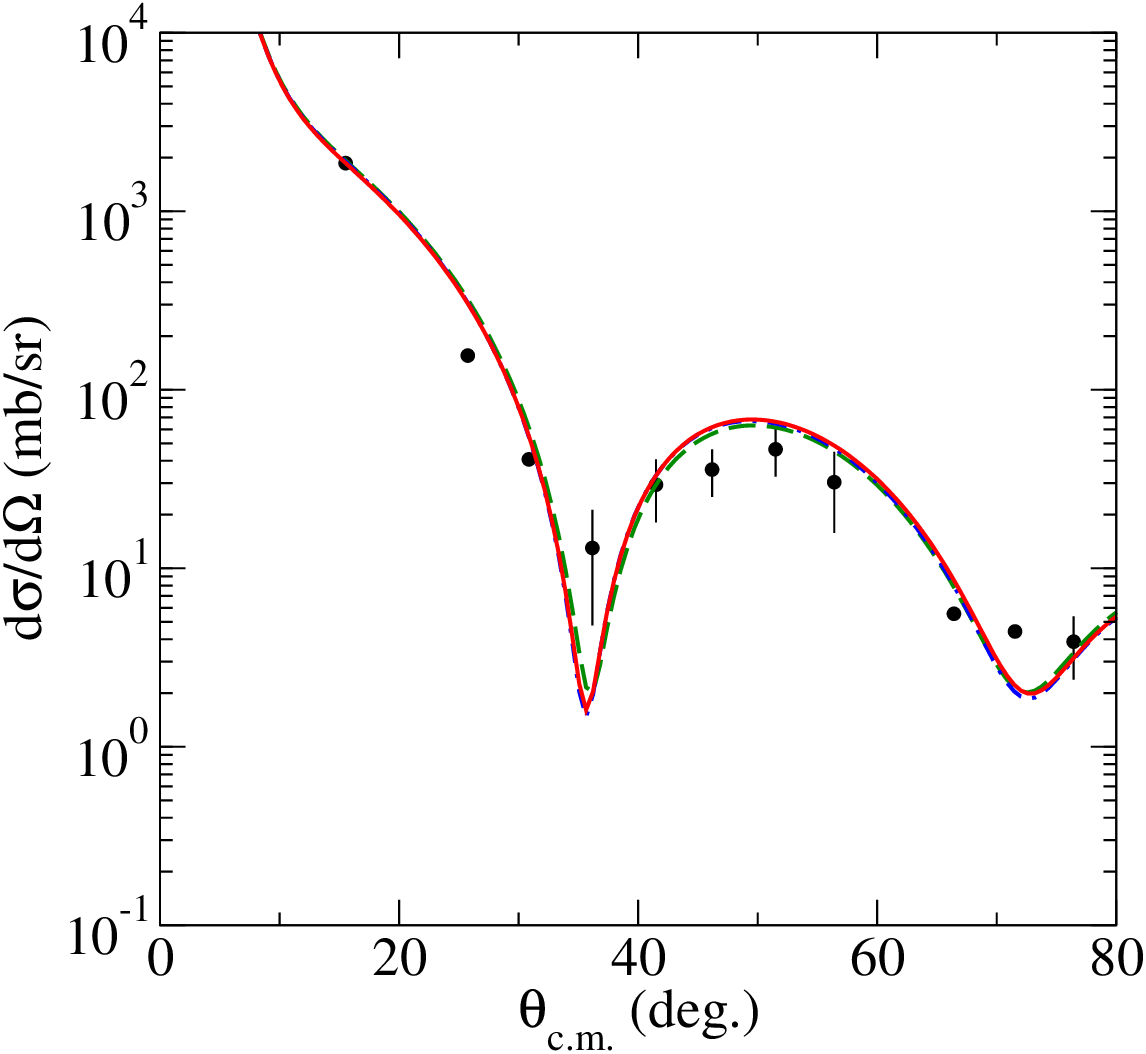}}
\caption{\label{s34-scatt} (Color online.) As for Fig.~\ref{s32-scatt}
  but for the elastic scattering of 29.8~MeV protons from $^{34}$S.}
\end{figure}

The results of the calculations from the three models for the
scattering of 29.8~MeV protons from $^{34}$S are compared to the data
\cite{Al85} in Fig.~\ref{s34-scatt}.
For $^{34}$S, there is very little difference between the results
obtained using the shell model (the same oscillator length was used as
for $^{32}$S) or either of the SHF models. All results are in quite reasonable
agreement with the available data \cite{Al85}. This is consistent with
the results for the scattering at 29.6~MeV for $^{32}$S.

\begin{figure}
\scalebox{0.5}{\includegraphics*{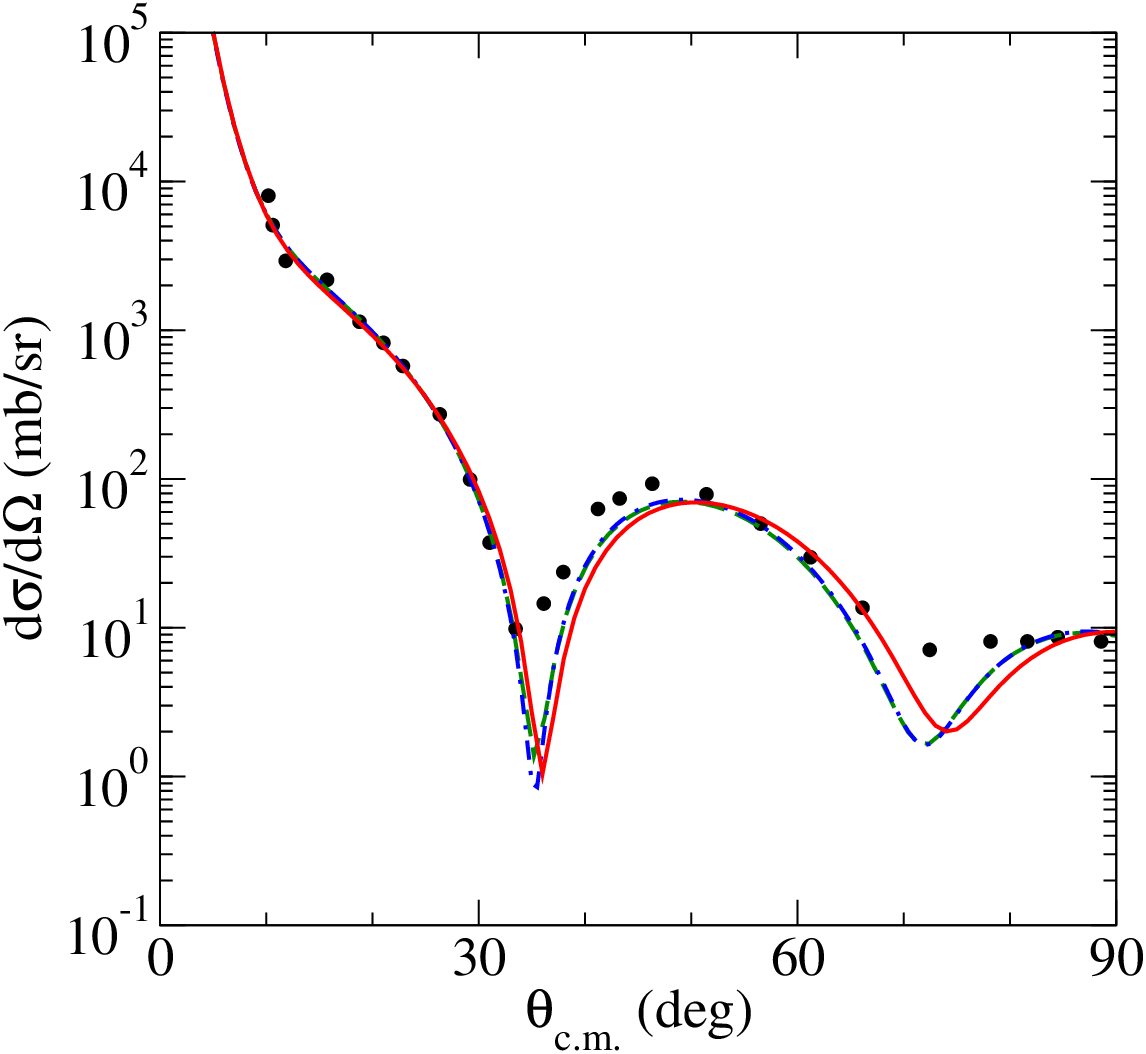}}
\caption{\label{s36-scatt} (Color online.) As for
  Fig.~\ref{s32-scatt}, but for the elastic scattering of 28~MeV
  protons from $^{36}$S.}
\end{figure}

Fig.~\ref{s36-scatt} shows the differential cross section for the
elastic scattering of 28~MeV protons from $^{36}$S.
Therein, the results of the calculations made using the shell ($b = 1.9$~fm), SHF,
and SHF-SM models are compared to the data \cite{Ho90}. All models
do equally well in describing the data, although the level of
agreement between the results and the data above $35^{\circ}$ is
somewhat poorer. Considering the problems in specifying the
microscopic optical potentials below 30~MeV \cite{Am00}, the level of
agreement is still reasonable. A previous JLM analysis \cite{Ma99} was
able to reproduce the data, but only by renormalising the
real and imaginary parts of the potential. Also, in that case, the
isovector part of the JLM potential was increased by a factor of 2.0.
No such renormalisations were required in the potential for the
results in the present work.

\begin{figure}
\scalebox{0.5}{\includegraphics*{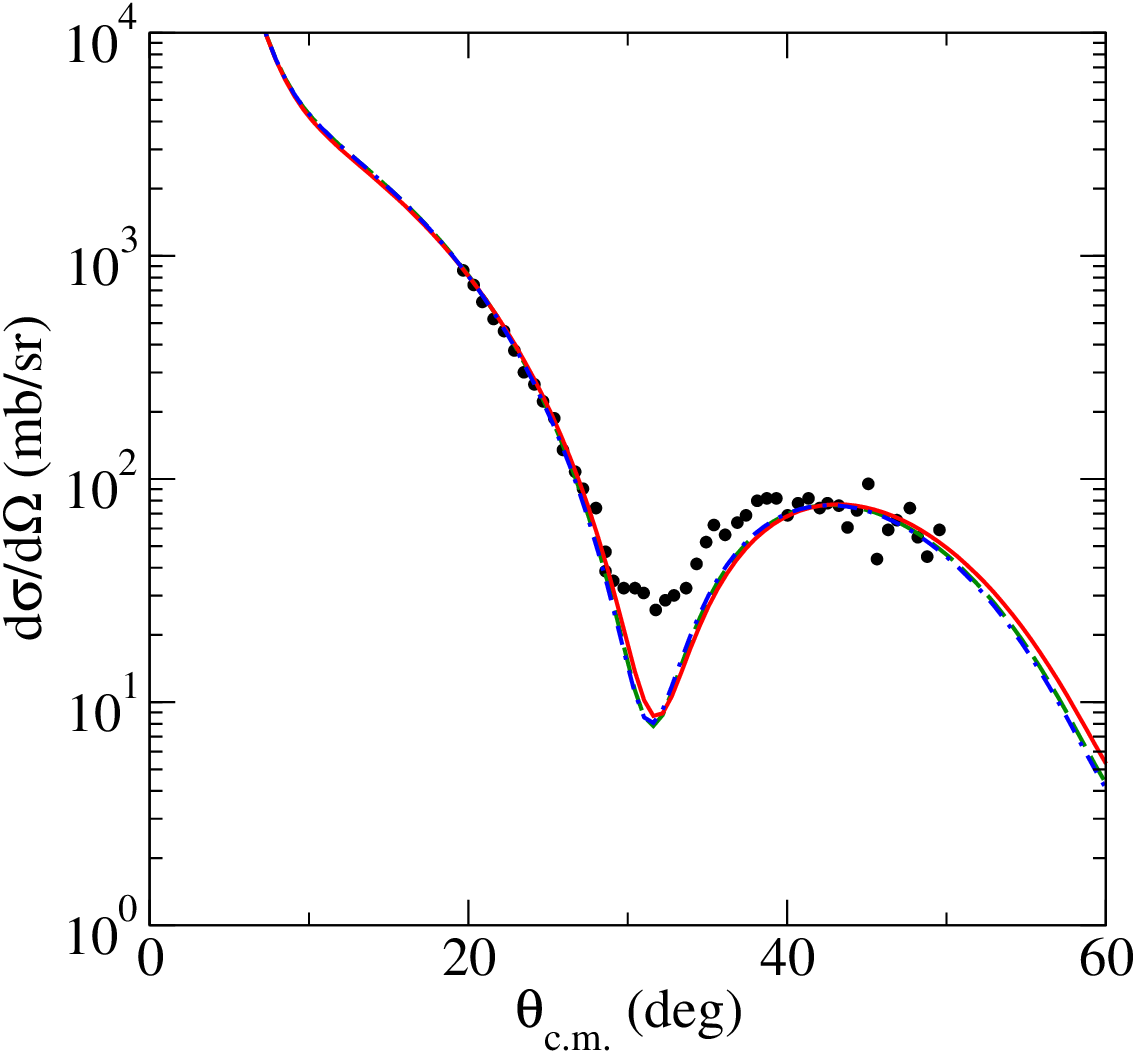}}
\caption{\label{s38-scatt} (Color online.) As for
  Fig.~\ref{s32-scatt}, but for the elastic scattering of 39~MeV
  protons from $^{38}$S. The shell model used in this case is a packed
  model.}
\end{figure}

The differential cross section for the elastic scattering of 39~MeV
protons from $^{38}$S is shown in Fig.~\ref{s38-scatt}.
Therein, the three models ($b = 1.9$~fm for the packed shell model) are
compared to the data \cite{Ke97}. All the models do equally well in
describing the data up to $30^{\circ}$. All the models underestimate the data in the region of the
minimum, up to $40^\circ$. The previous JLM model result \cite{Ma99} required a
larger renormalisation of the imaginary part of the potential compared
to the real part, which was not in keeping with the normalisations that model required
for the lighter isotopes.

\begin{figure}
\scalebox{0.5}{\includegraphics*{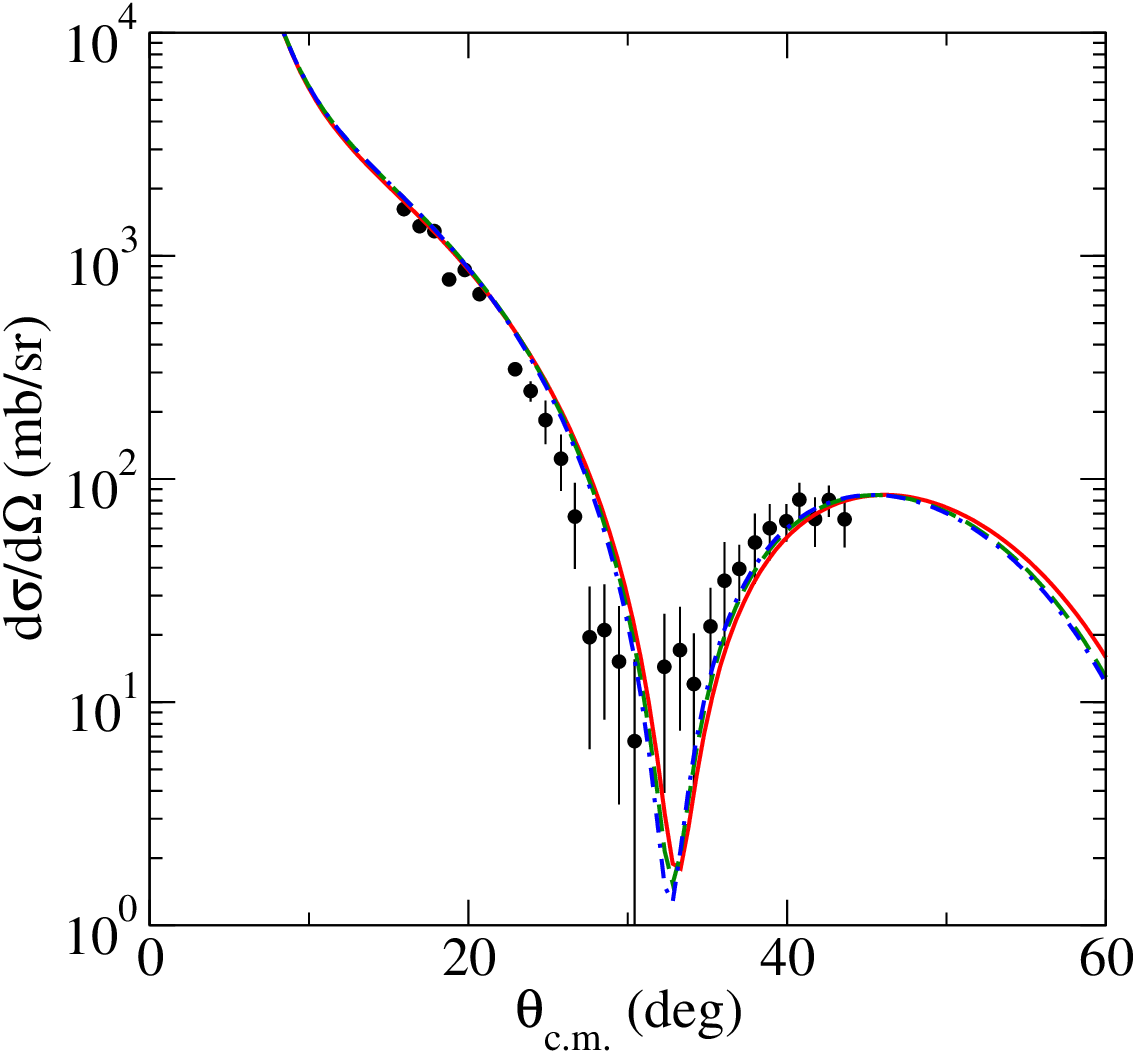}}
\caption{\label{s40-scatt} (Color online.) As for
  Fig.~\ref{s38-scatt}, but for the elastic scattering of 30~MeV
  protons from $^{40}$S.}
\end{figure}

Fig.~\ref{s40-scatt} shows the differential cross section for 30~MeV
elastic scattering of protons from $^{40}$S.
Therein, the data \cite{Ma99} are compared to the results from the
packed shell ($b = 1.9$~fm), SHF, and SHF-SM models. All the results
compare equally well with the data, and once more no renormalisations in the
potentials were necessary in achieving these results. However, the level of agreement is
difficult to gauge given the large error bars in the data.

%
%
\subsection{The Ar isotopes}
\begin{figure}
\scalebox{0.5}{\includegraphics*{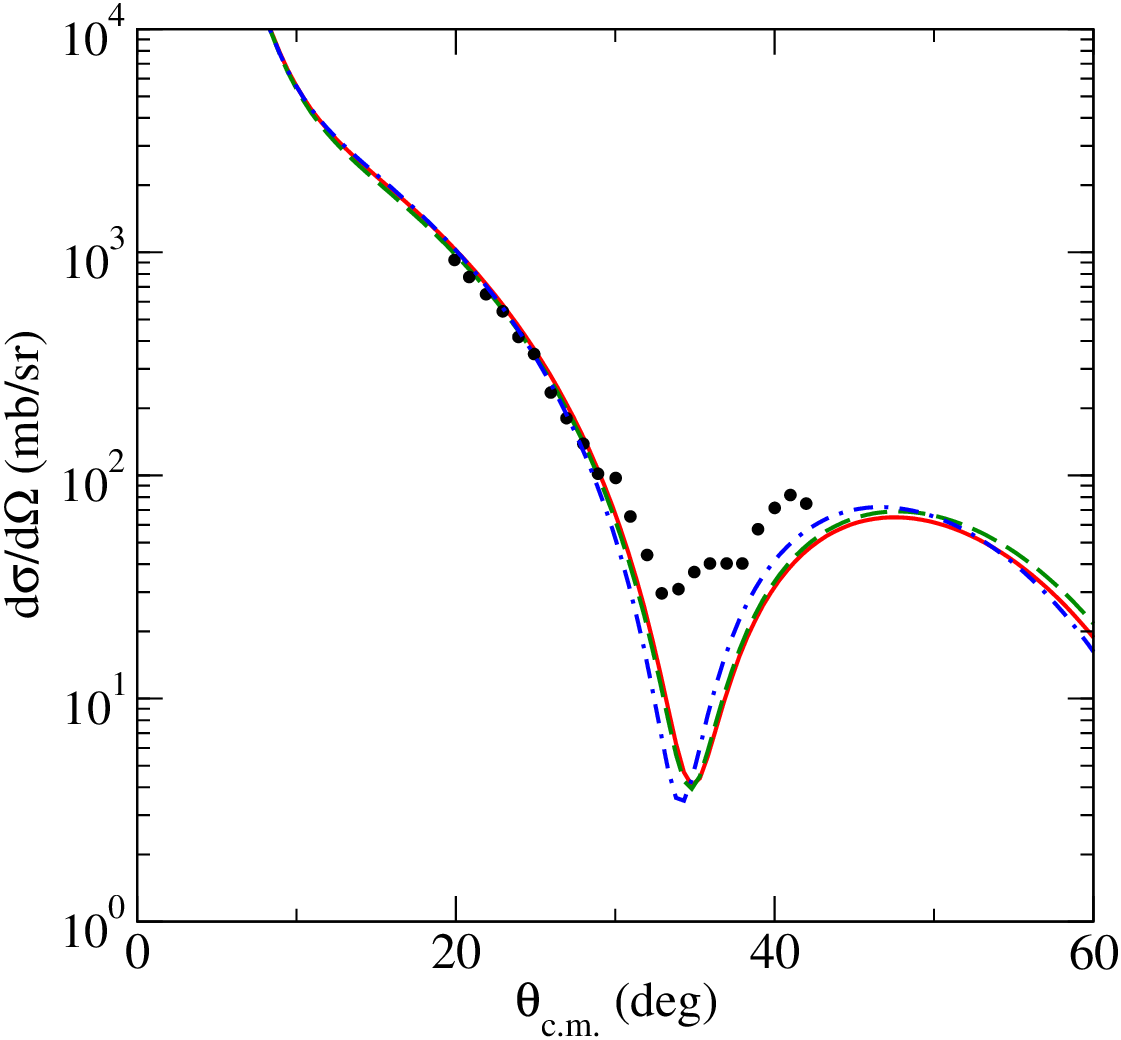}}
\caption{\label{ar36-scatt} (Color online.) Differential cross section
  for the elastic scattering of 33~MeV protons from $^{36}$Ar. The
  data \cite{Sc00} are compared to the results of the calculations
  made using wave functions obtained from the shell, the
  SHF, and the SHF-SM models.}
\end{figure}

We now turn our attention to the Ar isotopes, beginning with
$^{36}$Ar. Fig.~\ref{ar36-scatt} shows the comparison
of the data to the results of the calculations made using the wave
functions from the shell ($b = 1.85$~fm), basic SHF, and SHF-SM
models. All model results agree with the data up to $32^{\circ}$
quite well. However, beyond that the model results do not reproduce
either the shape or the magnitude of the data, although the SHF-SM
result does marginally better than the
other two.  An earlier microscopic (JLM model) analysis \cite{Sc00}
required independent renormalisations of
the real and imaginary parts of the effective $NA$ interaction of 0.94
and 0.92, respectively, to fit the data. One cannot conclude
as to whether the reproduction of the data is due to the underlying
description of the nucleus, or the renormalisations of the potential.
Also, those data beyond $30^{\circ}$ do not define a sharp diffraction
minimum, even one that may be smoothed by experimental
resolution. Another measurement may be required to explain the
anomalies between the models' predictions and to confirm the data.

\begin{figure}
\scalebox{0.5}{\includegraphics*{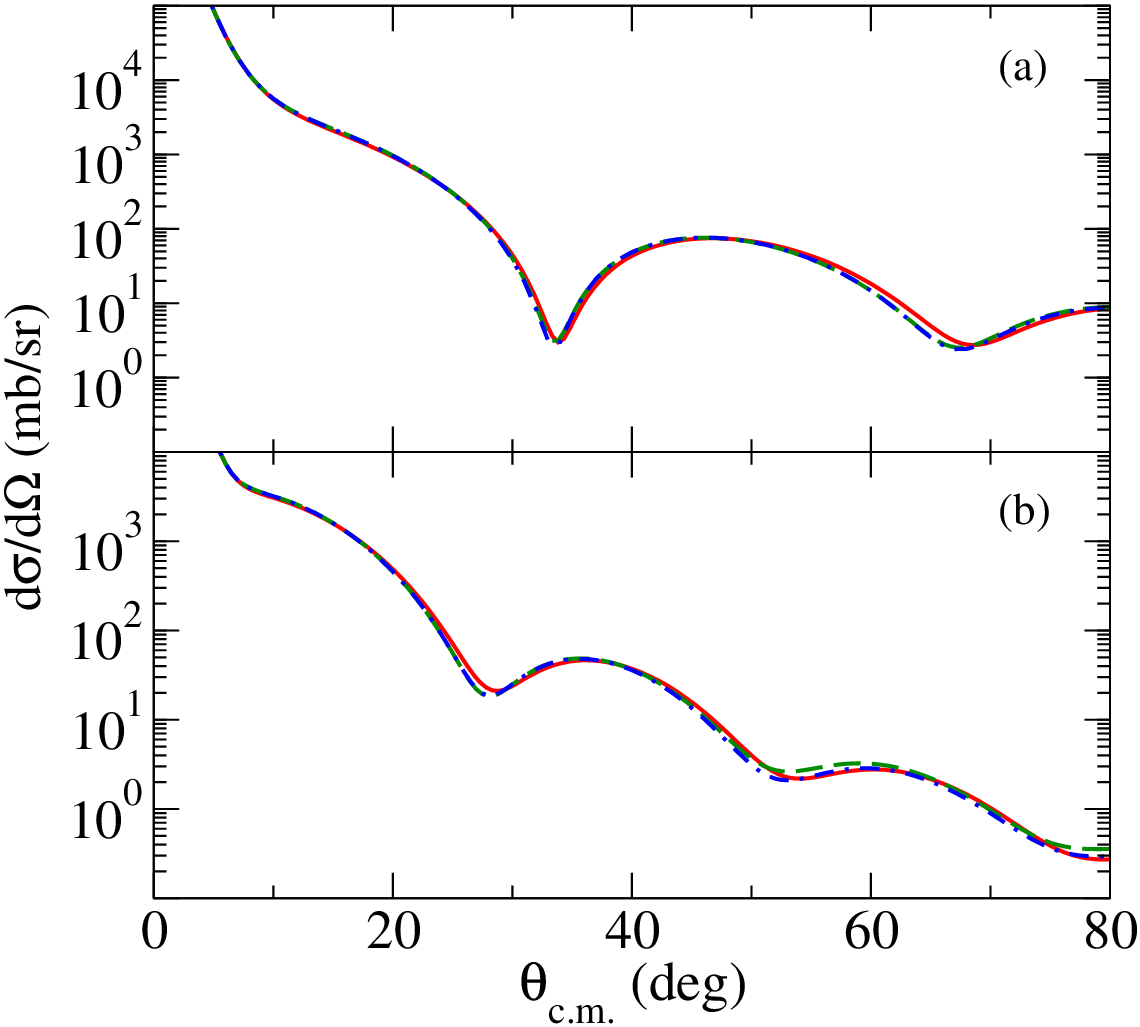}}
\caption{\label{ar38-scatt} (Color online.) Comparison of the
  differential cross sections obtained from calculations made using
  the shell, SHF, and SHF-SM models for the elastic scattering of 33~MeV (a) and 65~MeV (b)
  protons from $^{38}$Ar.}
\end{figure}

Fig.~\ref{ar38-scatt} presents the results of calculations made for
the elastic scattering of 33 and 65~MeV protons from $^{38}$Ar using
wave functions from the three models, where $b = 1.85$~fm is used for the
shell model densities.
There is very little difference between the three models.

\begin{figure}
\scalebox{0.6}{\includegraphics*{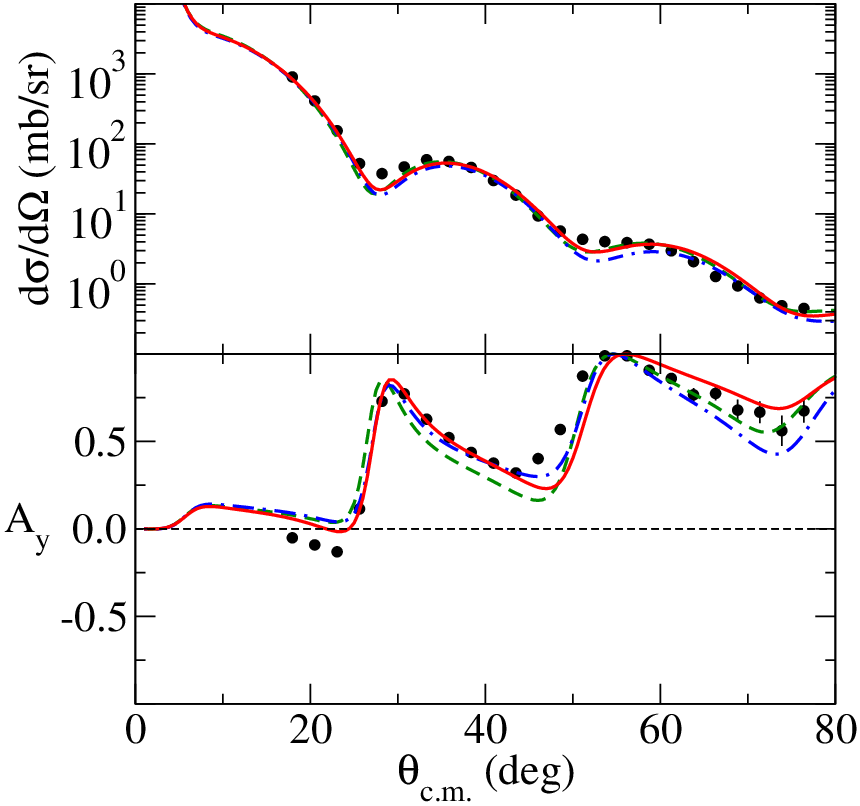}}
\caption{\label{ar40-scatt} (Color online.) Differential cross section
  and analysing power for the elastic scattering of 65~MeV protons
  from $^{40}$Ar. The data \cite{Sa82} are compared to the
  calculations made using a packed shell, the SHF, and the SHF-SM
  models.}
\end{figure}

The differential cross sections and analysing powers for the elastic
scattering of 65~MeV protons from $^{40}$Ar are displayed in
Fig.~\ref{ar40-scatt}.
Therein, the differential cross-section and analysing power data
\cite{Sa82} are compared to the results of the calculations made using
a simple packed shell model ($b = 1.85$~fm), the SHF, and SHF-SM
models. Of the three results, the best agreement with the data
comes from using the simple packed model. The differential cross section is
well reproduced also by the SHF model, while the SHF-SM model does
marginally less well. For the analysing power, all three models explain
the data reasonably well, though the best agreement is obtained with the
packed shell model.

\begin{figure}
\scalebox{0.5}{\includegraphics*{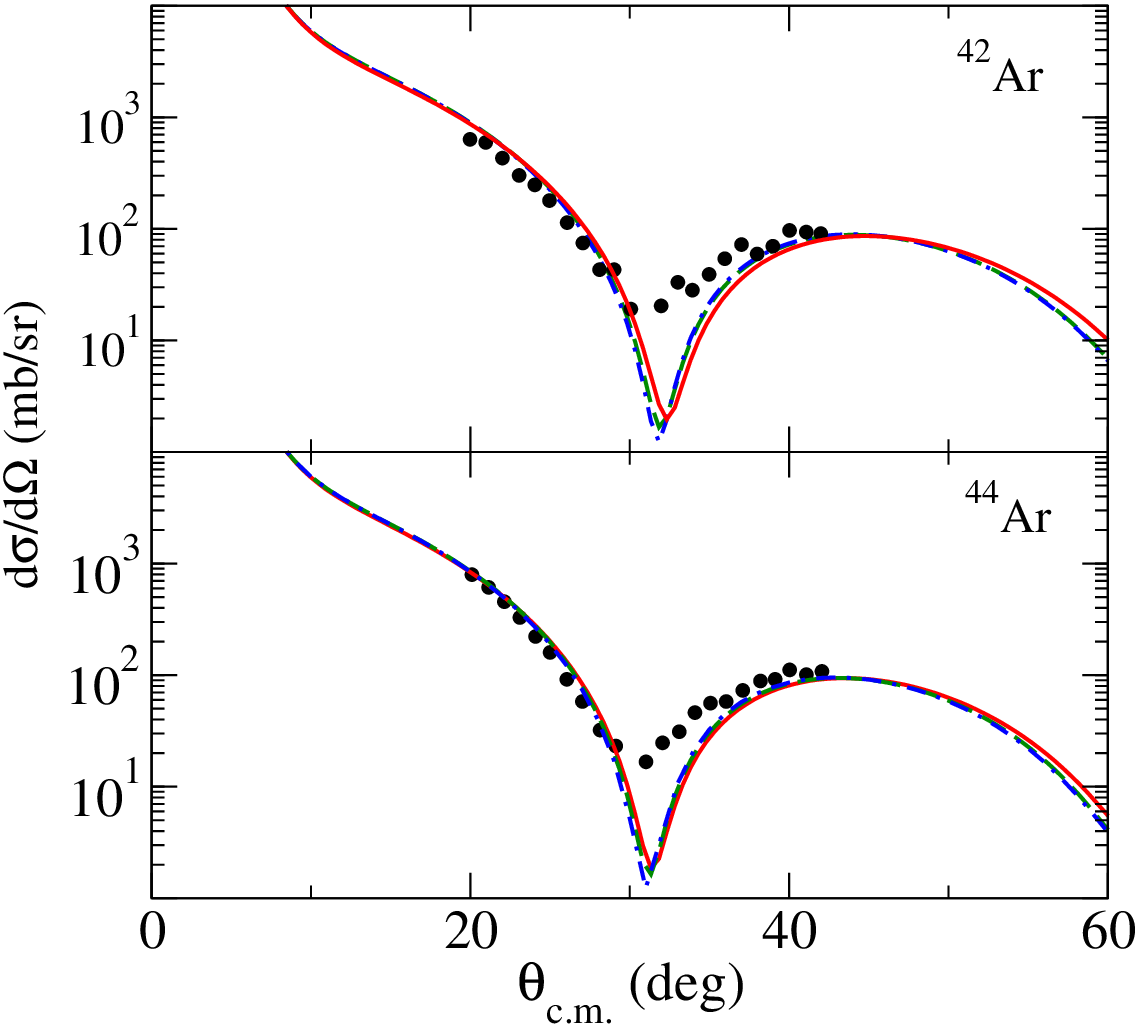}}
\caption{\label{ar42_44-scatt} (Color online.) Differential cross
  sections for the scattering of 33~MeV protons from $^{42}$Ar and
  $^{44}$Ar. The data \cite{Sc00} are compared to the results of the
  calculations from the packed shell, SHF,
  and SHF-SM models.}
\end{figure}

Fig.~\ref{ar42_44-scatt} shows the differential cross sections for the
33~MeV elastic scattering of protons from $^{42}$Ar and $^{44}$Ar.
Therein, the data \cite{Sc00} are compared to the results of the
calculations made using the simple packed ($b = 1.85$~fm and 1.88~fm
for $^{42}$Ar and $^{44}$Ar, respectively), the SHF, and SHF-SM
models. For both nuclei, there is no difference between the SHF
and SHF-SM results. For $^{36,42,44}$Ar, the models all give good results for
scattering to $\sim 30^{\circ}$, and underestimate the cross section
at larger angles. While the discrepancy is not as severe, and there is some
agreement between the model results and the data at larger angles in
magnitude and shape, we note that these data are from the same
experiment as that for $^{36}$Ar. The JLM analysis presented in the
earlier work \cite{Sc00} required the same renormalisations of the
real and imaginary parts. There is consistency
in those JLM analyses but the question remains as to whether the level of
agreement is due to the underlying structure or the renormalisations
themselves.

%
%
\subsection{The Ca isotopes}
\begin{figure}
\scalebox{0.55}{\includegraphics*{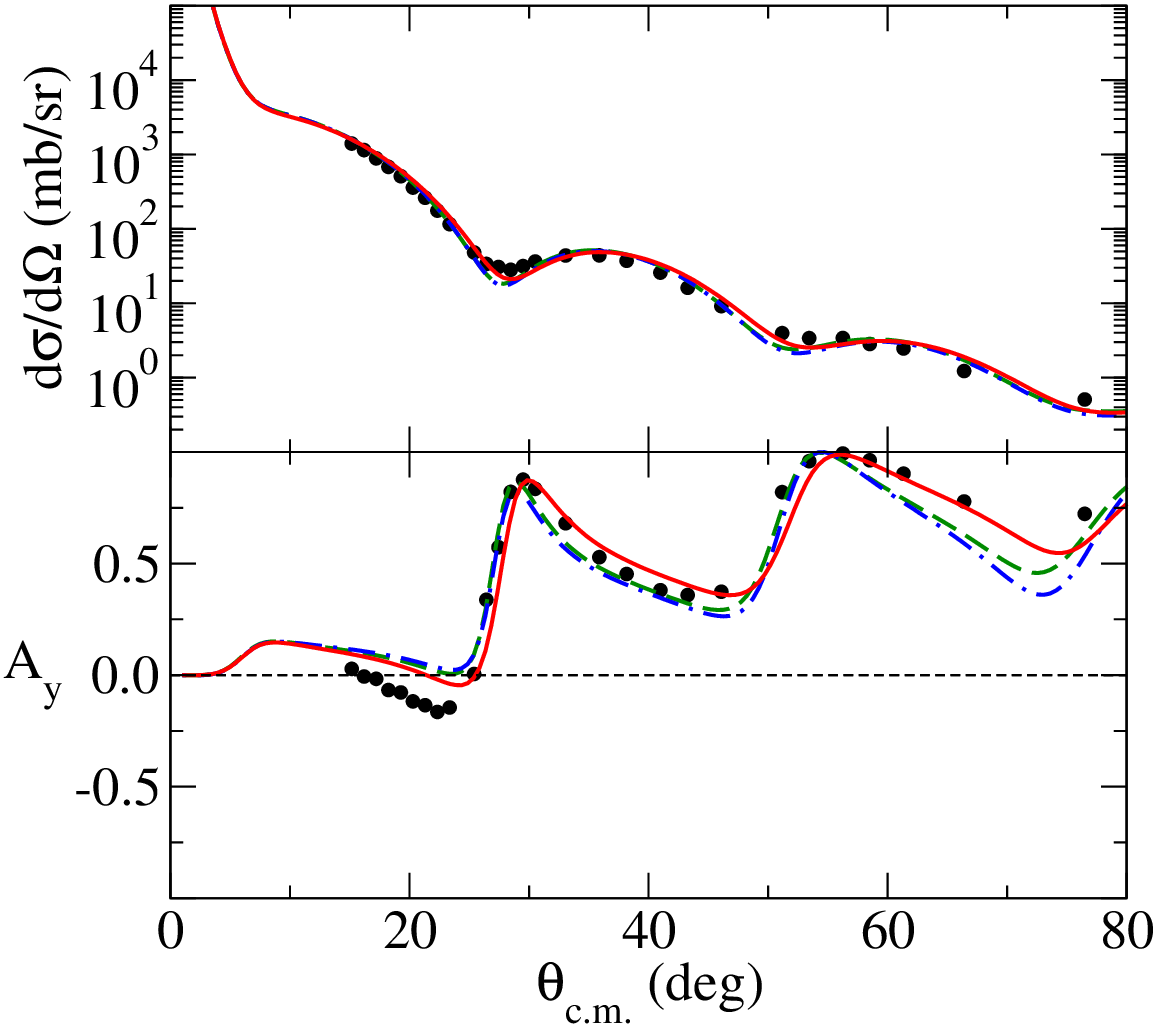}}
\caption{\label{ca40-scatt} (Color online.) Differential cross section
  and analysing power for the elastic scattering of 65~MeV protons
  from $^{40}$Ca. The data \cite{Sa82} are compared to the results of
  the calculations performed using a packed shell, SHF, and SHF-SM
  models.}
\end{figure}

The differential cross section and analysing power for the elastic
scattering of 65~MeV protons from $^{40}$Ca are displayed in
Fig.~\ref{ca40-scatt}.
Therein, the data \cite{Sa82} are compared to the results of
calculations made using the packed shell ($b = 1.9$~fm), the SHF, and
SHF-SM models. All models do equally well in describing the
differential cross-section data to $80^{\circ}$. It is in the
analysing power that differences emerge, with the packed shell model
providing a better description of the data.

\begin{figure}
\scalebox{0.68}{\includegraphics*{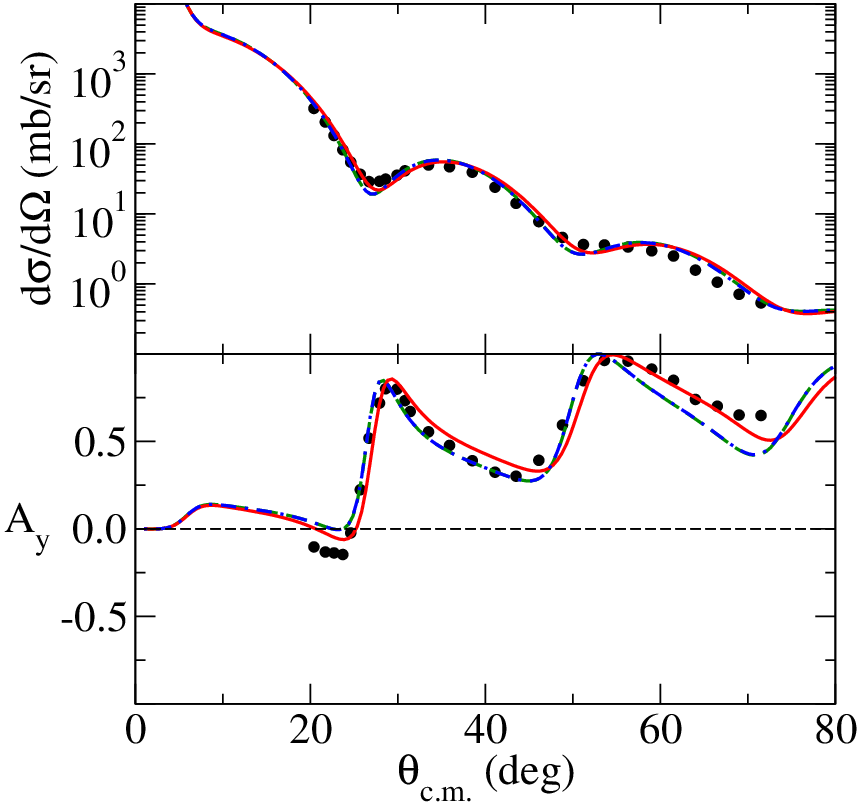}}
\caption{\label{ca42-scatt} (Color online.) Differential cross section
  and analysing power for the elastic scattering of 65~MeV protons
  from $^{42}$Ca. The data \cite{No81} are compared to the results of
  the packed shell model, SHF, and SHF-SM calculations.}
\end{figure}

Fig.~\ref{ca42-scatt} shows the differential cross section for the
elastic scattering of 65~MeV protons from $^{42}$Ca.
The data \cite{No81} for the differential cross section are well
described by the shell ($b = 1.9$~fm) and both SHF models, as was the case for scattering from
$^{40}$Ca. As with $^{40}$Ca, the packed shell model does better at
describing the analysing power. Note that the results from the SHF models are identical, as the
ground-state wave function is largely $( 0f_{\frac{7}{2}} )^2_\nu$, as indicated by the
occupation numbers.

\begin{figure}
\scalebox{0.5}{\includegraphics*{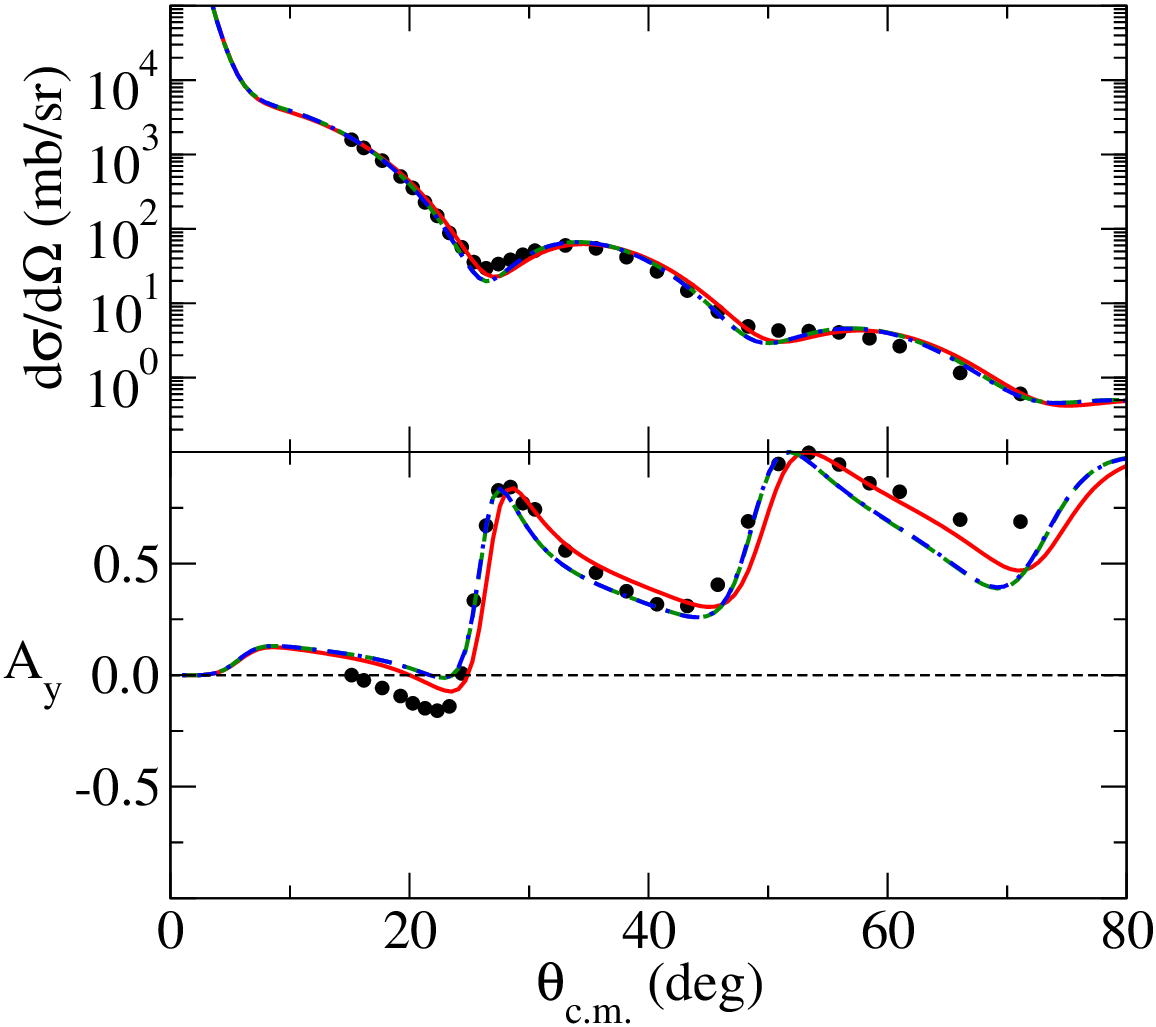}}
\caption{\label{ca44-scatt} (Color online.) Differential cross section
  and analysing power for the elastic scattering of 65~MeV protons
  from $^{44}$Ca. The data \cite{Sa82} are compared to the results of
  the calculations from the packed shell, SHF, and SHF-SM models.}
\end{figure}

A similar observation may be reached for the comparison of the results
of the calculations made from the packed shell ($b = 1.9$~fm) and SHF models with
data for the elastic scattering of 65~MeV protons from $^{44}$Ca, for
which the differential cross section and analysing power are displayed
in Fig.~\ref{ca44-scatt}.
Therein, the data \cite{Sa82} are compared to the results of the three
model calculations. The results for the differential cross section
compare quite well with the data, while the analysing power data are
better described by the packed shell model result. As with $^{42}$Ca, both
SHF models give identical results, as the ground state is predominantly $( 0f_{\frac{7}{2}} )^4_\nu$.

\begin{figure}
\scalebox{0.5}{\includegraphics*{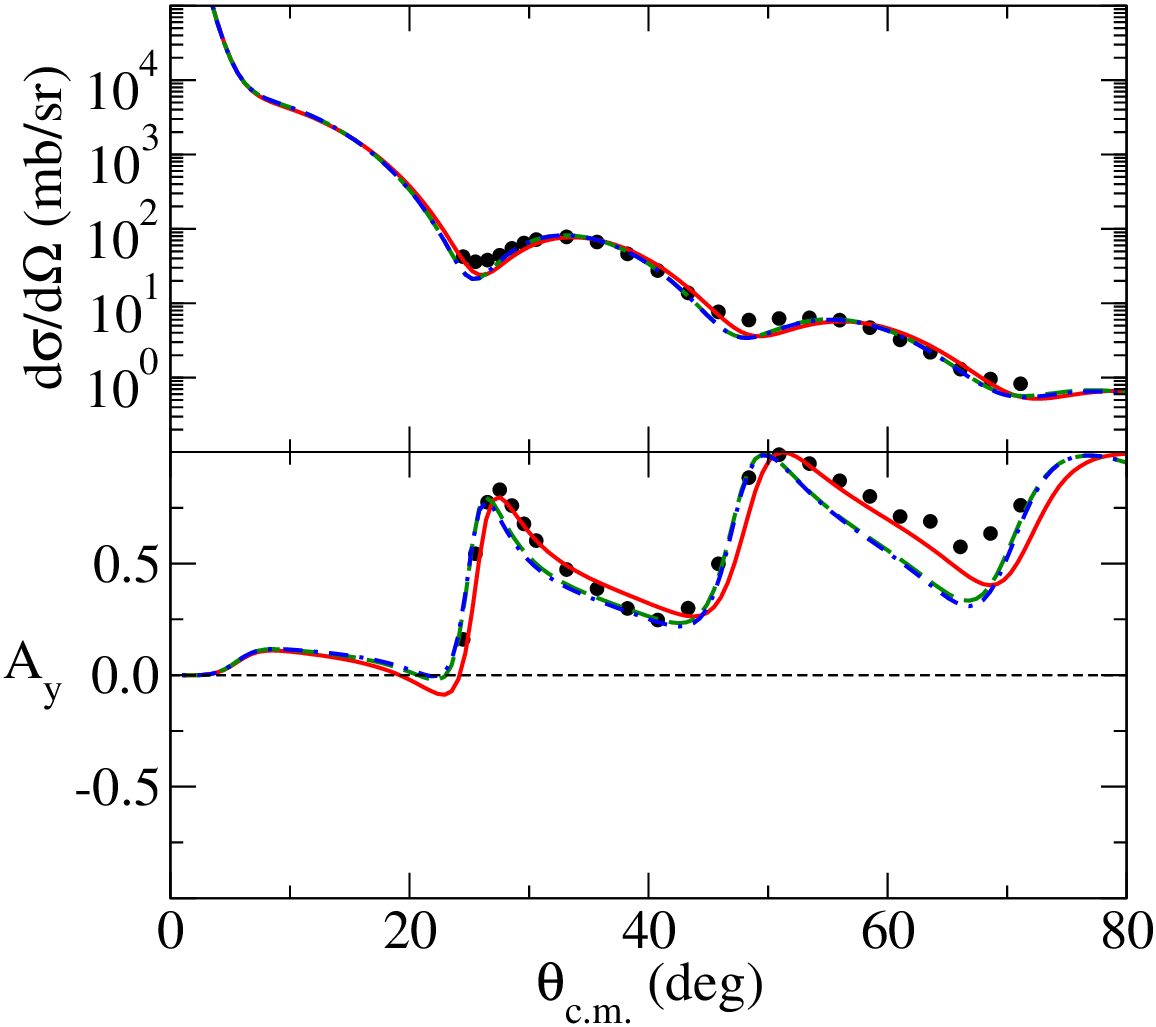}}
\caption{\label{ca48-scatt} (Color online.) As for
  Fig.~\ref{ca40-scatt} but for the scattering from $^{48}$Ca.}
\end{figure}

$^{48}$Ca provides for an interesting test of the SHF models, as that
nucleus is very well described by a filled $0f_{\frac{7}{2}}$ neutron
shell. Fig.~\ref{ca48-scatt} displays the differential cross section
and analysing power for
65~MeV elastic proton scattering from $^{48}$Ca. The data \cite{Sa82}
are compared to the results of the calculations made using all three
models, where the shell model is a closed $0f_{\frac{7}{2}}$ neutron
orbit ($b = 1.9$~fm). All three results of the model calculations
compare very well with the data. Yet, the SHF and SHF-SM models are
almost in complete agreement: the SHF assumes a closed
$0f_{\frac{7}{2}}$ neutron orbit, while the occupancies used in
constraining the SHF-SM, listed in Table~\ref{table-occ}, indicate
some mixing in $^{48}$Ca. The degree of that mixing is small, and that
is evident in the nearly identical predictions made between the SHF
and SHF-SM models, for both the differential cross section and
analysing power.

\begin{figure}
\scalebox{0.5}{\includegraphics*{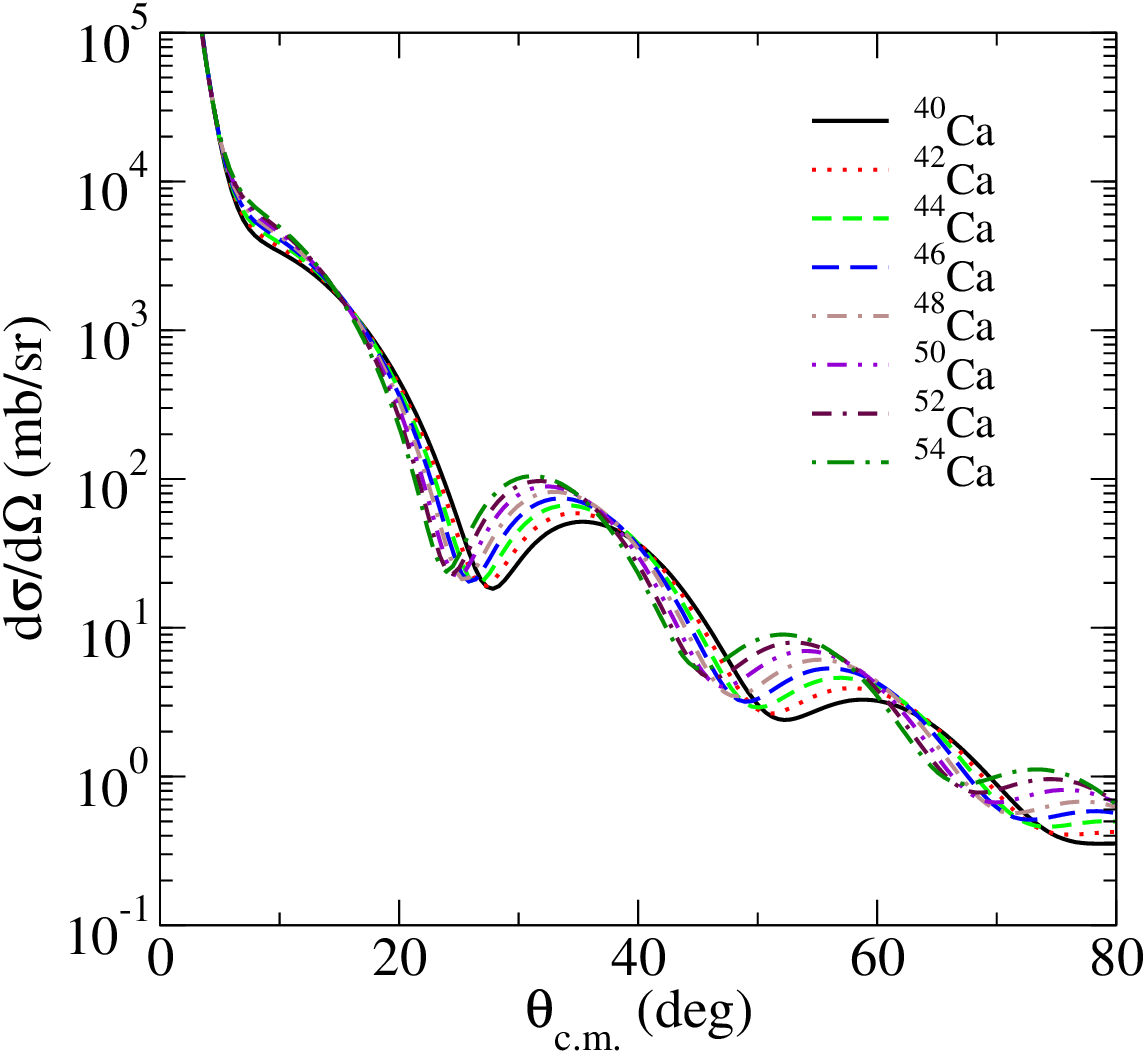}}
\caption{\label{ca-isotopes} (Color online.) Differential cross
  sections for the elastic scattering of 65~MeV protons from all
  even-even Ca isotopes from $^{40}$Ca to $^{54}$Ca. The SHF model was
  used to obtain the cross sections.}
\end{figure}

Fig.~\ref{ca-isotopes} displays the differential cross sections for
the elastic scattering of 65~MeV protons from all even-even Ca
isotopes from $^{40}$Ca to $^{54}$Ca. Therein, the SHF model was used
to obtain the differential cross sections.
As shown in the figure, there is a clear mass dependence in the
differential cross section, as expected. Whether the cross sections
for the heaviest Ca isotopes may be measured remains an academic
question as it depends on the availability of suitable Ca beams. It is
hoped that such beams will become available, as there is a question of
the presence of a magic number either at $^{52}$Ca or $^{54}$Ca
\cite{Ho02}. As the spin-orbit force may change close to the drip
line, it has been suggested \cite{Ho02} that the $0f_{\frac{5}{2}}$
orbit falls below the $1p_{\frac{3}{2}}$, in which case $^{52}$Ca
would no longer be doubly-magic.

\section{Conclusions}

We have presented results of calculations of elastic scattering of
protons from isotopes of S, Ar, and Ca, using models based on
Skyrme-Hartree-Fock methods. Two such models were used: one, the
standard SHF, while the other constrained the Hartree-Fock iterations
by specifying the nucleon occupancies, obtained from the shell model,
\textit{a priori}. For comparison, the shell model was also used to
calculate the scattering observables.

Comparisons of results from calculations of proton elastic scattering
from $^{28}$Si indicated that using densities obtained from either the
SHF models was as good, if not better, than using those from the shell
model. Of the two SHF models, the SHF-SM model was found to be
slightly better.

That the SHF-SM model gives a better description of the nuclear
density was evident for $^{32}$S since the differential cross
section and analysing power were better described using that
model structure. However, for the heavier isotopes of sulfur, that was 
not the case. Using either model for the structure, equivalent descriptions of the elastic scattering data
were obtained. The shell model constraint on the
Hartree-Fock procedure did not improve the level of agreement. It is
of note that the descriptions of the elastic scattering worsened for
large angles in these cases; serious renormalisations were required of
the JLM potentials to obtain fits to the data. As the
specification of the Melbourne potential does not allow for such
renormalisations, we hope that more data may be collected to confirm
the data used herein. The same conclusions were reached for the Ar
isotopes, for which more data are also needed.

The Ca isotopes provided a sensitive control on the specification of
the SHF densities, as $^{48}$Ca is very well described by a filled
$0f_{\frac{7}{2}}$ neutron shell. In that case, one expects the
SHF and SHF-SM results to be equal. This was shown to be so in the
analyses of the differential cross-section and analysing-power data
for the elastic scattering. Also, we considered the mass dependence on
the scattering from $^{40}$Ca to $^{54}$Ca and found that dependence
to be smooth. As there is interest in the descriptions of the
structures of $^{52}$Ca and $^{54}$Ca, we hope that scattering data
will be measured in the future when beams of those two exotic nuclei
become available.

Together with the reaction cross section, measurements of the
differential cross sections will elicit much information on the
structures of these neutron-rich nuclei. With the increase in the
number of radioactive beam facilities it is hoped that not only will
there be data taken for new nuclei, but also of those already
available to provide valuable confirmation of previous experiments.

\begin{acknowledgments}
  This research was supported by grants from the Australian Research
  Council and the National Research Foundation of South Africa.  The authors thank
  B. A. Brown for supplying some of the occupation numbers used in this work. W.~R.
  also would like to thank the School of Physics, University of
  Melbourne, and Rhodes University for hospitality and support.
\end{acknowledgments}


\bibliography{isotopes_shf}

\end{document}